Spin-driven tidal pumping: Tidally driven changes in planetary spin coupled with secular interactions between planets

by Richard Greenberg[1], Christa Van Laerhoven[1], and Rory Barnes[2]

[1]Lunar and Planetary Laboratory, University of Arizona

[2]Astronomy Department, University of Washington

Authors' address:
Lunar and Planetary Laboratory
University of Arizona
1629 East University Blvd.
Tucson, Arizona 85721-0092 USA

Corresponding author:
Richard Greenberg
Email: greenberg@lpl.arizona.edu
Phone: 1-520-621-6940
Fax: 1-520-621-9692

**Abstract**

In a multiplanet system, tides acting on the inner planet can significantly affect the orbital evolution of the entire system. While tides usually damp eccentricities, a novel mechanism identified by Correia et al. (2012) tends to raise eccentricities as a result of the tides' effect on the inner planet's rotation. Our analytical description of this spin-driven tidal (SDT) effect shows that, while the inner planet's eccentricity undergoes pumping, the process is more completely described by an exchange of strength between the two eigenmodes of the dynamical system. Our analysis allows derivation of criteria for two-planet coplanar systems where the SDT effect can reverse tidal damping, and may preclude the effect's being significant for realistic systems. For the specific case quantified by Correia et al., the effect is strong because of the large adopted tidal time lag, which may not be appropriate for the assumed Saturn-like inner planet. On the other hand, the effective $Q$ for any given planet in exotic circumstances is very uncertain, so the SDT effect could play a role in planetary evolution.

# 1. Introduction

The orbital eccentricities *e* of planets close to their stars are damped by tides raised on the star and on the planet (*e.g.* Jeffreys 1961, Goldreich and Soter 1966, Kaula 1968, Jackson et al. 2008). The tides also affect semi-major axes *a*, whose changes can modify *e*-damping rates on comparable timescales (Jackson et al. 2008). Evolution becomes more complex and interesting in multiplanet systems, where angular momentum is periodically exchanged among the planets through secular interactions (*e.g.* Rodrìguez et al. 2011). The periodic changes in the eccentricities can be described by a sum of eigenmodes that affect all the planets (*e.g.* Brouwer and Clemence 1961, Murray and Dermott 1999). As a result, tidal effects on one planet can damp all the planets' *e* values. Moreover, tidal changes in that planet's *a* also modify the eccentricities (Wu and Goldreich 2002, Greenberg & Van Laerhoven 2011, Laskar et al. 2012).

Recently an additional way that tides may modify eccentricities in a multiplanet system has been identified by Correia et al. (2012, hereafter CBL), involving the role of planetary rotation, as follows: The secular interactions among planets are partially controlled by any effect that contributes to the precession of the major axis of any planet, *e.g.* General Relativity (GR) and the oblateness of the star and planet. The planet's oblateness depends on its spin rate, and tidal dissipation drives the spin toward a rate that depends on the orbital eccentricity. Because the secular interactions (exchanging angular momentum) among the planets cause periodic changes in eccentricities, the spin rate of a close-in planet may change periodically. The latter process is dissipative, as it depends on distortion of the figure of the planet, so the spin rate (and the corresponding oblateness) must lag relative to the secular change in e. Hence the secular behavior is out of phase with a periodic change in the oblateness that affects it, resulting in a pumping of eccentricity.

CBL demonstrated this effect for a specific hypothetical system. They based their model on equations for secular interactions among planets valid for large, as well as small, eccentricities (e.g. Mardling 2007), using a common formulation for the change in spin rate due to tides. These equations are not generally amenable to an analytic solution, but numerical integration demonstrated the potential magnitude of this effect. Further, CBL derived an analytic solution with the assumption of a very low amplitude of oscillations, which showed how the tidal phase lag can drive the pumping. These results were surprising, both because tides are generally expected to damp eccentricities and because of the large magnitude of the computed pumping effect.

Here we demonstrate that similar results can be obtained with an analytical solution based on lower-order classical secular theory. This approach has the advantage of being readily generalizable, allowing for an assessment of how the effect depends on the parameters of a system. It also helps elucidate various aspects of the physics of the process. For example, while the effect has been described with an emphasis on the pumping of one planet's *e*, what is really happening is an exchange of strength between eigenmodes (and an exchange of angular momentum between the planets), such that while one planet's orbit becomes more eccentric, the other is circularized. For this reason, rather than emphasizing the pumping alone, we refer to this process as the "spin-driven tidal (SDT) effect".

Our analytic solution involves several simplifying approximations, which are commensurate with the great uncertainty in actual tidal parameters and internal dissipation processes. For the specific case studied by CBL, our results are similar, despite the different suite of approximations. For the sake of the comparison, we used the same tidal parameters as CBL; however the dissipative time-lag assumed for illustrative purposes in CBL's hypothetical case may not be consistent with solar-system constraints (**Section 4.1**).

## 2. Analytic solution

### 2.1 Relevant formulae from secular theory

The secular interactions in a co-planar two-planet system can be described by

$$de_1/dt = A_{12}\, e_2 \sin(\varpi_1 - \varpi_2) \tag{1a}$$

$$de_2/dt = A_{21}\, e_1 \sin(\varpi_2 - \varpi_1) \tag{1b}$$

$$d\varpi_1/dt = A_{11} + A_{12}\, (e_2/e_1) \cos(\varpi_1 - \varpi_2) \tag{1c}$$

$$d\varpi_2/dt = A_{22} + A_{21}\, (e_1/e_2) \cos(\varpi_2 - \varpi_1) \tag{1d}$$

where $\varpi$ is the longitude of pericenter. The values $A_{ij}$ are well-known functions of the masses of the planets relative to the star's, the mean motions, and the semimajor axes (*e.g.* Brouwer and Clemence 1961, Murray and Dermott 1999). Here we have included terms in the disturbing functions through second-order in the eccentricities; a discussion of this approximation is in **Section 3.2.1**. The coefficients $A_{11}$ and $A_{22}$ represent the precession rates due to several effects: the $e=0$ component of the other planet; GR; the planet's tidal elongation aligned with the direction of the star; and the oblateness of the planet and star; (See Raggozine and Wolf (2009) for a discussion of the relative importance of these terms.) The pumping effect discovered by CBL comes from the dependence of $A_{11}$ on the oblateness of the inner planet.

The usual definitions $h = e \sin \varpi$ and $k = e \cos \varpi$ yield the well-known linear equations

$$dh_1/dt = A_{11}k_1 + A_{12}k_2 \tag{2a}$$

$$dk_1/dt = -A_{11}h_1 - A_{12}h_2 \tag{2b}$$

$$dh_2/dt = A_{21}k_1 + A_{22}k_2 \tag{2c}$$

$$dk_2/dt = -A_{21}h_1 - A_{22}h_2 \tag{2d}$$

The solution of these equations is also well-known (e.g. Brouwer and Clemence 1961, Murray and Dermott 1999); we briefly summarize the key features here. In the solution, each eccentricity vector (h,k), with magnitude e and direction $\varpi$, is a sum of two vectors corresponding to the eigenmodes of the solution as shown in **Figure 1**. The eigenfrequencies are

$$g_{p,m} = (A_{11} + A_{22} \pm S) / 2 \tag{3}$$

where

$$S = \sqrt{(A_{11} - A_{22})^2 + 4A_{12}A_{21}} \tag{4}$$

Following the notation of Wu and Goldreich (2002), we designate the modes as m or p depending on the sign before the square root in Eq. (3) being minus or plus, respectively. The eigenvectors of the solution give the ratios of the eccentricity vector components $F_p = e_{1p}/e_{2p}$ and $F_m = e_{1m}/e_{2m}$ as:

$$F_p = e_{1p}/e_{2p} = -(A_{11} - A_{22} + S)/2A_{21} \tag{5a}$$

$$F_m = e_{1m}/e_{2m} = (A_{11} - A_{22} - S)/2A_{21} \tag{5b}$$

As shown in **Figure 1**, note that the directions for $e_{1m}$ and $e_{2m}$ are aligned, whereas those for $e_{1p}$ and $e_{2p}$ are anti-aligned. The magnitudes of these constants are given by the initial conditions.

The eccentricity vector components rotate at a rate given by the corresponding eigenfrequency, i.e. the angles $\alpha$ and $\beta$ defined in **Figure 1** increase at the rates $g_m$ and $g_p$, respectively. The difference between these directions is $\theta = \beta - \alpha$. $\theta$ increases at the constant rate $g_p - g_m$, which equals $S$ according to Eq. (3).

By the law of cosines,

$$e_1^2 = e_{1m}^2 + e_{1p}^2 + 2\, e_{1m}\, e_{1p} \cos\theta . \tag{6}$$

If we define $t = 0$ such that $\theta = 90°$ at that time, and define $e_0^2 = e_{1m}^2 + e_{1p}^2$ and $E^2 = 2e_{1m}e_{1p}$, we have

$$e_1^2 = e_0^2 + E^2 \sin St \tag{7}$$

The SDT effect is based on variations in $A_{11}$, specifically those that result from changes in the spin rate of the inner planet (which change its oblateness, and hence change the orbital precession rate $A_{11}$). The changes in $A_{11}$ are small over the secular timescale $2\pi/S$. Thus, the analytic method developed by Greenberg and Van Laerhoven (2011) for the case where slow changes in $a_1$ gradually modify all $A_{ij}$ values can readily be modified to apply to the case of a gradual change in $A_{11}$ only. We simply replace $\delta a_1$ with $\delta A_{11}$ in Greenberg and Van Laerhoven's Eqs. (13a) and (13b), yielding

$$\delta e_{1p} = F'_p\, \delta A_{11}\, e_{2p} + F_p(a_1)\, \delta e_{2p} \tag{8a}$$

$$\delta e_{1m} = F'_m \, \delta A_{11} \, e_{2m} + F_m(a_1) \, \delta e_{2m} \tag{8b}$$

where $\delta a_1$ has been replaced by $\delta A_{11}$ and $F'$ for each mode is now the partial derivative of $F$ with respect to $A_{11}$ rather than with respect to $a_1$, so $F'_m$ and $F'_p$ can be found by differentiating Eqs. (5):

$$F'_p = -(1 + (A_{11} - A_{22}) / S) / 2A_{21} = +F_p / S \tag{9a}$$

$$F'_m = -(1 - (A_{11} - A_{22}) / S) / 2A_{21} = -F_m / S \tag{9b}$$

From Eq. (15) of Greenberg and Van Laerhoven we now have

$$de_{2p}/dt = (A_{21}/S)(e_{2m} \, F'_m \cos\theta + e_{2p} \, F'_p) \, dA_{11}/dt \tag{10a}$$

$$de_{2m}/dt = (A_{21}/S)(e_{2p} \, F'_p \cos\theta + e_{2m} \, F'_m) \, dA_{11}/dt \tag{10b}$$

(Note we have corrected typographical errors in the published version of these equations by removing the minus signs and restoring missing prime signs.) Inserting these equations into Eqs. (8) above, we also have

$$de_{1p}/dt = F'_p \, e_{2p} \, dA_{11}/dt + F_p(a_1) \, de_{2p}/dt \tag{11a}$$

$$de_{1m}/dt = F'_m \, e_{2m} \, dA_{11}/dt + F_m(a_1) \, de_{2m}/dt \tag{11b}$$

Thus we have expressions for the expected change in all the components of the eccentricities as a result of gradual changes in $A_{11}$.

## 2.2 Changes in spin rate

Next we evaluate the effect on $A_{11}$ of the tidal changes in the inner planet's rotation. For a body in a circular orbit, tidal dissipation should lead to synchronous rotation with no further dissipation. However, with an eccentric orbit, there may be a tidal torque even during synchronous rotation (when $\omega = n$) because the orbital angular velocity is variable over each orbit. Thus one expects the spin rate $\omega$ to vary as a function of $e$ and $\omega$, with an equilibrium rate other than synchronous if $e \neq 0$ (e.g. Greenberg and Weidenschilling 1984, Efroimsky 2012, Makarov and Efroimsky 2013, Ferraz-Mello 2013).

CBL adopted a specific model of the tidal torque as a function of the rotation rate $\omega$ and eccentricity under the influence of the tidal torque, based on a widely used assumption that the dissipative lag of the tidal response is a constant time delay $\Delta t$, independent of the driving frequencies. Although the realism of that assumption is questionable (see **Section 3.2.1**), for purposes of comparison we use the same expression for the change in the rotation rate $\omega$:

$$d\omega/dt = A\left[(1 + \tfrac{27}{2}e^2) - (1 + \tfrac{15}{2}e^2)\omega/n\right] \tag{12}$$

where

$$A \equiv n^2 \frac{3k}{\xi Q} \frac{M}{m} (R/a)^3 \tag{13}$$

and where $n$ is the orbital mean motion, $k$ is the second-order Love number (usually called $k_2$), $\xi$ is the coefficient of the moment of inertia of the planet (0.4 for a uniform sphere), $Q$ is the tidal dissipation parameter of the planet, $M$ is the mass of the star, $m$ is the mass of the planet, and $R$ is its radius. As noted by CBL, $Q = 1/(n \Delta t)$. For simplicity, where planetary parameters have no subscript below, they refer to the inner planet.

According to Eq. (12), at any instant the spin rate $\omega$ always changes toward a value for which $d\omega/dt$ would be zero, which is $\omega_e = n(1 + 6e^2)$, ignoring terms of fourth-order in $e$. If the eccentricity undergoes periodic change, as it does during secular interactions, $\omega$ chases toward an ever-changing value $\omega_e$. We define $\omega_0$ as the value of $\omega_e$ for the average value of $e^2$ over a secular cycle, i.e. $\omega_0 = n(1 + 6e_0^2)$, and define $\varepsilon = \omega - \omega_0$. In order to incorporate the secular changes in $e$, we insert the expression from Eq. (7) into the rate of change of the spin, yielding

$$d\varepsilon/dt = A (6E^2 \sin St - \varepsilon/n) \tag{14}$$

which has the solution

$$\varepsilon = -6E^2 n \left[ \frac{\delta}{(1+\delta^2)} \right] \cos St + 6E^2 n \left[ \frac{1}{(1+\delta^2)} \right] \sin St \tag{15}$$

where $\delta = Sn/A$.

The cosine term in Eq. (15) represents the part of this spin response that is out of phase with the secular variation of $e^2$ in Eq. (7). According to CBL, and as confirmed below, this out-of-phase component is crucial for the tidal pumping. The phase lag of $\varepsilon$ relative to $e^2$ depends on the value of the ratio $Sn/A$. As defined in Eq. (13), $A$ represents the tidal dissipation in the inner planet. For very small $A$, i.e. little dissipation, $\delta$ is large, so the cosine term is small, but the sine term is much smaller; the variation of $\varepsilon$ lags by nearly ¼ cycle (90°) behind the secular variation of $e$. At the opposite extreme, with $A \gg Sn$, the sine term dominates and there is very little phase lag. The reason is that, according to Eq. (12), with large enough $A$ the spin rate would change easily, allowing it to keep up with the change in $e$. So, if there is too much dissipation, the crucial phase lag is diminished and one might expect the SDT effect to be weak. The effect is strongest when $\delta = 1$ and the phase lag is 45°. This point is also addressed by Correia (2011) and Correia et al. (2013).

The variation in spin rate given by Eq. (15) affects the secular behavior by modifying the coefficient $A_{11}$. Recall that any effect that tends to change in the rate of precession of the pericenter gives a change in $A_{11}$. The oblateness of the planet has the following effect on the pericenter longitude (c.f. CBL's Eq. 8):

$$d\varpi/dt = n\,(k/2)\,(M/m)\,(R/a)^5\,(\omega/n)^2 \qquad (16)$$

or using the definition of ε

$$d\varpi/dt = n\,(k/2)\,(M/m)\,(R/a)^5\,(\omega_0^2 + 2\omega_0\varepsilon)/n^2 \qquad (17)$$

Because n and a vary slowly compared with the timescale for secular variations, in effect the only non-constant quantity on the right hand side of Eq. (17) is ε. Eq. (17) represents the contribution of the planet's oblateness to the precession rate $A_{11}$. Therefore $dA_{11}/dt$ is the derivative of Eq. (17).

$$dA_{11}/dt = k\,(M/m)\,(R/a)^5\,(\omega_0/n)\,d\varepsilon/dt \qquad (18)$$

Differentiating the expression for ε from Eq. (15) and inserting it into (18) yields

$$dA_{11}/dt = 6\,E^2\,nS\,k\,(M/m)\,(R/a)^5\,(\omega_0/n)\,\{[\delta/(1+\delta^2)]\sin St + [1/(1+\delta^2)]\cos St\} \qquad (19)$$

Next we insert $dA_{11}/dt$ from Eq. (19) into Eq. (10a) for $de_{2p}/dt$, noting that $\cos\theta = \sin St$ according to Eqs. (6) and (7). We also average $de_{2p}/dt$ over the secular cycle (St from 0 to 2π). Only the terms of Eq. (10a) and (19) that contain sin St contribute to a non-zero average value. Thus,

$$de_{2p}/dt = 3\,E^2\,e_{2m}\,n\,k\,(M/m)\,(R/a)^5\,(A_{21}F'_m)\,(\omega_0/n)\,[\delta/(1+\delta^2)]. \qquad (20)$$

Inserting the definition of E and $F'_m$ (from Eq. 9b) yields

$$de_{2p}/dt = 3\,e_{1m}\,e_{1p}\,e_{2m}\,n\,k\,(M/m)\,(R/a)^5\,[(S - A_{11} + A_{22})/S]\,(\omega_0/n)\,[\delta/(1+\delta^2)] \qquad (21a)$$

Similarly, Eqs. (10b), (11a), and (11b) become the following:

$$de_{2m}/dt = (e_{2p}/e_{2m})\,[(A_{11} - A_{22} + S)/(A_{11} - A_{22} - S)]\,de_{2p}/dt \qquad (21b)$$

$$de_{1p}/dt = (e_{1p}/e_{2p})\,de_{2p}/dt \qquad (21c)$$

$$de_{1m}/dt = (e_{1m}/e_{2m})\,de_{2m}/dt \qquad (21d)$$

**2.4 Evaluation of the rates of change**

These expressions can be evaluated and compared with the numerical results of CBL, who considered two similar hypothetical systems. The one they considered in most detail had the following properties: a stellar mass of 1.05 solar masses; an inner planet with $m = 0.18\,M_J$, Saturn's radius, $k = 0.5$, and $\xi = 0.2$; and an outer planet with $m_2 = 0.2\,M_J$. Initial orbits had $a = 0.25$ AU (implying $n = 51$ rad/yr) and $a_2 = 1.8$ AU. Those values define the $A_{ij}$ matrix, and hence the eigenmodes, of the system. Note that $A_{11}$ is dominated by the effect of the outer planet, as well as the contribution to the precession of the inner planet by GR. The eigenmodes give us the frequency of the secular cycle $S = 1.96\times10^{-5}$ rad/yr.

The initial conditions for the secular interaction adopted by CBL had $e_1 = 0.3$, $e_2 = 0.4$, and $\varpi_1 - \varpi_2 = 180°$, leading to the following values of the eccentricity components: $e_{1p} = 0.367$; $e_{1m} = 0.067$; $e_{2p} = 0.022$; $e_{2m} = 0.378$. **Figure 2** illustrates the eccentricity vectors to scale. The fact that each planet is dominated by a different mode (planet 1 by mode p, and planet 2 by mode m) indicates that the planets are relatively uncoupled dynamically, although still coupled enough to demonstrate the SDT effect. (As discussed in **Section 3.2.2** below, the weak coupling is implicitly assumed in the analytical formulation derived by CBL.)

Next we evaluate the SDT effect for this hypothetical system, for comparison with the results of CBL. For the tidal parameter $Q$, CBL used the formula $1/Q = n_1\Delta t$ where $\Delta t$ is the lag time of the tidal response of the figure of the planet. They assumed that the tidal lag $\Delta t$ was independent of the driving frequency with a value $\Delta t = 200$ sec. For the comparison, we use the same value, although we address the uncertainty regarding this parameter in **Section 4.1**.

Substituting all the parameters for CBL's hypothetical system into our solution (Eqs. 21) yields

$de_{1p}/dt = \phantom{-}0.0050$/Gyr
$de_{1m}/dt = -0.0003$/Gyr
$de_{2p}/dt = \phantom{-}0.0003$/Gyr
$de_{2m}/dt = -0.0016$/Gyr

These results show that the most rapid change is in the eccentricity components that dominate $e_1$ and $e_2$ (c.f. **Figure 2**), that is $e_{1p}$ and $e_{2m}$, respectively. The inner planet's eccentricity $e_1$ would increase by about 1% in a billion years at this rate, while $e_2$ decreases. The solution shows that, while the SDT effect pumps one eccentricity, the other one is damped. Rather than simply pumping, the SDT effect involves an exchange between the two eigenmodes, as discussed in Section 3.3. So $e_2$ decreases because it is dominated by the opposite mode from the one that dominates $e_1$.

These values can be compared with the rates of change of eccentricities obtained by CBL for this case. They graphically presented (their Fig. 1b) results obtained in two ways: a numerical integration of the relevant differential equations and an analytic solution that assumes small amplitudes of the secular oscillations of $e_1$ and $e_2$ (equivalent to small $e_{1m}$ and $e_{2p}$). The rates for the eccentricity variation displayed there are considerably faster than given by our solution above, a difference attributable to the large eccentricity values for this case. In the next section, we discuss this comparison in more detail.

**3. Comparison of solutions**

**3.1 Values from CBL's numerical integration**

The rates of change we evaluated in **Section 2.4** assumed the same hypothetical system considered by CBL, which allows for a direct comparison of results of the two methods of solution. Because CBL solved the governing equations numerically, they did not solve for eigenmodes as we have. Nevertheless, the behavior plotted by CBL is qualitatively consistent with the eigenmode solution, so approximate values of the eigenmode components and their evolution can be inferred from their plots.

Our eigenmode solution (*e.g.* **Figure 2**) shows that both $e_1$ and $e_2$ oscillate with a small amplitude about a substantial mean value: $e_1$ oscillates about $e_{1p}$ with amplitude $e_{1m}$; $e_2$ oscillates about $e_{2m}$ with amplitude $e_{2p}$. Similarly, CBL's Figure 1b shows $e_1$ and $e_2$ each oscillating with an amplitude much smaller than its mean value. Thus, the results are qualitatively similar and we can read values from their figure at time t=0 that correspond to our solution: $e_{1p}$ = 0.4; $e_{1m}$ = 0.1; $e_{2p}$ = 0.03; $e_{2m}$ = 0.37. Those values are in reasonably good agreement with our analytical solution of the secular behavior to lowest order in eccentricities: $e_{1p}$ = 0.367; $e_{1m}$ = 0.067; $e_{2p}$ = 0.022; $e_{2m}$ = 0.378.

The slopes of curves in CBL's Figure 1b give the rates of change of the eccentricity components at t=0. The mean values of $e_1$ and $e_2$ change by roughly 0.04/Gyr and −0.02/Gyr, equivalent to our $de_{1p}/dt$ = 0.005/Gyr and $de_{2m}/dt$ = −0.002/Gyr, respectively. The numerical integration thus gives SDT rates about ten times faster than our formulation. Similarly, the rate of change of the amplitudes of oscillation is an order of magnitude faster than our values of $de_{1m}/dt$ and $de_{2p}/dt$.

Such a discrepancy is not surprising, given that our analytic solution is based on secular theory valid only to second order in eccentricities, whereas CBL's hypothetical case involves *e* values as large as 0.5. In **Section 3.2.2**, we confirm that the magnitude of the discrepancy is as to be expected for a case with such a large eccentricities.

## 3.2 Analytic solutions

Analytic formulae for the SDT effect reveal its dependence on the parameters of any given system, as long as the assumptions involved in the derivation are met. In contrast, a numerical solution, while potentially more precise, only applies to the specific case that is evaluated. An analytic solution can thus serve as a tool for predicting the effect's significance for planetary systems of various configurations. Moreover, an analytic solution, as long as it gives a reasonably accurate description of the qualitative behavior, can reveal a great deal about the physical processes involved.

### 3.2.1 The small-eccentricity assumption

The disadvantage, of course, is that analytical approaches involve significant assumptions. Our solution depends on including mutual perturbations of the planets only through second order in eccentricities, so it can be accurate only if both eccentricities are small. Given that the hypothetical system of CBL involved large eccentricities, our solution cannot be a precise representation of the behavior. Nevertheless, even in this case it is accurate enough to show that the SDT effect is comparable to direct tidal damping.

Moreover, we have found previously that classical secular theory can give a reasonably good qualitative description of secular behavior and of longer-term orbital evolution, even with such large eccentricities (Van Laerhoven and Greenberg 2012). For example, for a study of tidal damping, Mardling (2007) had considered a hypothetical system with large eccentricities, and appropriately retained high-order terms in her governing equations, which then required a numerical solution. When we investigated the same system using our analytical approach, we obtained similar results (Van Laerhoven and Greenberg 2012). The solution elucidated the behavior in terms of the eigenmodes of the system, showing that it was qualitatively the same as the previously well-understood evolution for more nearly circular orbits.

While including eccentricities to a high order in a formulation would be ideal, that level of precision might not be comensurate with other uncertainties or approximations. For example both CBL and Mardling (2007) evaluated the $A_{ij}$ matrix using a quadrupole representation of the potential of one planet at the other, which is valid only to lowest order in $a_1/a_2$. Compared with the precise evaluation based on Laplace coefficients (*e.g.* Brouwer and Clemence 1961, Murray and Dermott 1999), that approximation introduces an error of about 2% for the hypothetical systems constructed in those studies. That level of error is acceptable, but it demonstrates that terms of high-order in eccentricity are not necessarily meaningful in comparison and may give a more-than-warranted impression of accuracy

A more significant issue regarding the validity of high-order terms in eccentricity stems from uncertainty in the tidal response of the figures of real planets. For example, the expressions for tidal effects on spin and orbital elements adopted by CBL are predicated on a widely used assumption that the tidal response lag is constant in time, independent of frequency. That assumption is attractive because expressions for high-order *e* terms based on it are readily available (e.g. Hut 1981). Yet there is no reason to believe that real planets behave in that way. For example, Ogilvie (2009) has shown that tidal dissipation in a Saturn-like planet may vary greatly, and irregularly, with small changes in frequency. Similarly, Efroimsky and Williams (2009) and Greenberg (2009) have argued that simplistic assumptions about effective values of *Q* and dependence on frequency may be very inaccurate, especially in the context of effects on planetary rotation (Makarov and Efroimsky 2013). The actual frequency dependence of tidal lag may be quite different from any common assumptions (Ferraz-Mello et al. 2009, Ferraz-Mello 2013). The low-order terms (such as those used in our analysis) are less sensitive to these uncertainties.

Therefore, while our analytic approach cannot claim great precision, it is appropriate given the great uncertainties of the problem, especially those regarding the value of *Q* or its frequency dependence. Moreover, our analysis is accurate enough to give insight into the mechanics of the processes involved, thus complementing other approaches.

### 3.2.2 Comparison of analytic approaches

As noted in **Section 1**, CBL also considered an analytic approximation (their Eq. 37) in addition to their numerical solution, which helped give insight into the phenomenon that they discovered. For this purpose, they assumed that the oscillations of $e_1$, of $e_2$ and of the rotation rate were small, so that they could linearize the equations about the mean values. That approach correctly revealed that the reason for the pumping was the phase lag of the spin response relative to the secular variation of the orbits.

However, the application of that formulation is limited. The requirement that the oscillation amplitudes must be near zero means either (a) both eigenmodes must be weak and thus both planets must have nearly circular orbits, or (b) the ratios $e_{1m}/e_{2m}$ and $e_{2p}/e_{1p}$ must both be very small. Those ratios are equivalent to the normalized eigenvectors, which are functions of the masses and semi-major axes. Condition (b) means that the planets must be only weakly coupled through secular interactions, because each planet's eccentricity is strongly dominated by a different eigenmode. **Figure 2** (as well as CBL's Figure 1) shows that this condition is only moderately well met for the case considered by CBL: $e_1$ oscillates about its mean value of 0.4 with a peak-to-peak amplitude of about 0.2. That the amplitude is even this small reflects the great separation between the planets in this case, where $a_2/a_1 > 7$. To have truly small-amplitude oscillations would require even weaker secular coupling. Therefore, the accuracy of CBL's analytic solution is limited to fairly special cases, with widely separated orbits or sufficiently small masses.

Given those limitations, it is striking that CBL's Figure 1b shows their analytical evaluation of the rate of increase of $e_1$ (from their Eq. 37) to be about 0.04/Gyr, which is very close to their numerical integration result. However, our own evaluation of CBL's Eq. 37, using either their mean values of $e_1$ and $e_2$ over the secular cycle or equivalently our evaluation of $e_{1p}$ and $e_{2m}$, gives a value over 0.08/Gyr. The value 0.04/Gyr would be obtained if we (inappropriately) used CBL's initial values of $e_1$ and $e_2$., which are not the average values over a secular cycle.

The evaluations of the two analytic approaches (CBL's and ours from **Section 2**) compared with the numerical integration of the controlling differential equations can be summarized as follows: The CBL analytic formula gives a rate about two times too large and ours gives a rate about ten times too small.

The difference between these results can be understood in terms of the different assumptions of the two analytic formulations. CBL's assumes that the amplitudes of secular oscillations are small, which implicitly requires that the planets are only weakly coupled. That condition roughly applies to their hypothetical system because of the very wide spacing between the planets. So in this case their formula gives a reasonable, order-of-magnitude result. On the other hand, our formulation is valid for even for strongly coupled systems, but requires that the eccentricities not be large. This restriction explains why our analytic evaluation gives much smaller rates, because the hypothetical system of CBL involves highly eccentric orbits. Indeed, inspection of the formulation by CBL (e.g. their Eq. 31) shows that the higher-order terms in $e$ have very large coefficients. It can be shown that for this system, those terms are dominant and can account for the order-of-magnitude difference between the results.

Thus, for such large-*e* cases, CBL's formulation is more accurate, if and only if the weak-interaction requirement is met. On the other hand, it is important to emphasize that the high-order e terms depend on the specific tidal model adopted for the derivation. CBL's analysis, as well as their numerical integration, depends on the assumption that the tidal parameter *Q* is inversely proportional to the frequency of tidal distortion, i.e. the time lag is constant. For any other tidal behavior, the higher-order terms in *e* in CBL's analytic solution may no longer be valid, and the solution loses its advantage. Because of the great uncertainty regarding the actual frequency dependence of the tidal response of planetary materials and structure, for eccentricities as large as those in this hypothetical system, no evaluation of the SDT effect analytic or by numerical integration can be quantitatively reliable.

For a system in which the assumptions of both analytic approaches are satisfied, i.e. where the eccentricities are not too large and the interactions are not too strong, our expression for $de_{1p}/dt$ (Eq. 21c above) and CBL's expression for $de_1/dt$ (their Eq. 37) are identical, as expected. One can confirm this identity by retaining only lowest-order-*e* terms in CBL's formula, and expressing their formula in our notation by comparing our Equations (1) with their equivalent differential equations (their 6, 7, and 8). Then their formula agrees perfectly with ours if and only if $4A_{12}A_{21}/(A_{11}-A_{22})^2 \ll 1$, in other words weak coupling. That requirement is equivalent to the low amplitude oscillations explicitly assumed by CBL. The two analytic converge to the same formula where both their sets of assumptions are satisfied.

### 3.3 Insights from our analytic solution

Several aspects of the SDT effect are revealed by the analysis in **Section 2**. One interesting result is the magnitude of the phase shift in the spin response relative to the eccentricity oscillation, which CBL showed is crucial to the process. With the value $\Delta t = 200$ sec as adopted by CBL, our tidal coefficient *A* (defined in Eq. 13) is $A = 1.66 \times 10^{-4}$ rad/yr$^2$. Thus, $\delta \equiv nS/A = 6.05$, not far from the value 1 that gives the critical cosine term in Eq. (15) its maximum possible value. If $\Delta t$ had been chosen to be only six times larger, the SDT effect would have reached a maximum; and if $\Delta t$ were even larger, the SDT effect would be reduced.

The phase lag is also unexpectedly large. CBL interpreted their results with the assumption that the phase lag of the spin relative to the secular oscillation was small, which seemed reasonable because tidal effects usually involve small lags. For a time lag in the tidal figure response of the planet of a few minutes (as assumed by CBL), and a period of secular variations of the orbits of $2\pi/S = 3 \times 10^5$ yr, giving a ratio of $\sim 10^{-11}$, one might expect the phase lag to be comparably small. However, the cosine term in Eq. (15) is much larger than the sine term; with $\delta = 6.05$, the phase lag of the spin relative to the secular driver is $\sim 80°$. Our analysis shows that the phase lag is actually quite large in this case.

Our solution also shows that the SDT effect is not simply an eccentricity-pumping process. In fact, while one planet's eccentricity does increase, the SDT effect actually exchanges amplitude between the two eigenmodes of the system. While the mode that dominates the inner planet increases (the "pumping"), the mode that dominates the outer planet decreases. In physical terms, the SDT effect transfers orbital angular momentum from the inner to the outer planet.

This result suggests a re-interpretation of the long-term evolution modeled by CBL. On the basis of their evaluation of the tidal pumping process, they numerically integrated the evolution of their hypothetical system over billions of years. Over ~3 Gyr, the mean value of $e_1$ and the oscillation amplitude of $e_2$ both increased, as we would expect with strengthening of mode p in our formulation. Similarly, the mean value of $e_2$ and the oscillation amplitude of $e_1$ both decrease in accord with the weakening of mode m.

In that calculation, ignoring the effect of tides on $a_1$, CBL found that eventually $e_2$ hit a value of zero, after which its oscillation amplitude began to decrease. This behavior is also consistent with the secular theory. Inspection of **Figure 2** shows that if mode p were to strengthen and mode m to weaken, eventually $e_{2p}$ and $e_{2m}$ will be equal, allowing $e_2$ to reach zero during the secular cycle. After that point in time, $\varpi_1 - \varpi_2$ librates about 180°, and the oscillation amplitude is controlled by mode m rather than mode p. This transition of $\varpi_1 - \varpi_2$ from circulation to libration is similar to what occurs if one (or both) of the planets experience eccentricity damping (Barnes and Greenberg 2006, Mardling 2007, Van Laerhoven and Greenberg 2012). It occurs in that case because the damping of one planet's eccentricity is shared between the planets, such that both eigenmodes damp. Usually (but not always) one mode damps more quickly than the other (Van Laerhoven & Greenberg 2012), so that when only one mode remains, the major axes are either aligned or anti-aligned. In principle, the same thing could happen if one or both eccentricities were pumped up (Chiang & Murray 2002), because generally one eigenmode would grow more quickly than the other, eventually becoming much larger than the other, so the major axes become aligned (c.f. Fig. 1). In the case of the SDT effect, one mode grows while the other shrinks, which also would eventually result in such alignment.

When CBL then repeated the simulation, now including the effect of tides on $a_1$, they found a different long-term evolution. After ~2 Gyr the evolution was no longer governed by the SDT effect, but rather both eccentricities (and in our formulation, both eigenmodes) began to decrease. They attributed this change to the decrease in $a_1$, which enhances the tidal damping of the eccentricities. However, that explanation is only part of the story for a couple of reasons. First, moving the inner planet closer to the star should also strengthen the SDT effect, at least partially offsetting the fact that it strengthens the direct tidal damping of $e_1$. More important is that the pumping of $e_1$ (*i.e.* the increase in $e_{1p}$) depends on the square of the amplitude of mode m (according to Eqs. 21a and 21c), and after 2 Gyr the SDT effect has reduced mode m significantly. In that way, eccentricity pumping by the SDT effect is self-limiting over time.

**4. Potential significance of the SDT effect**

**4.1 Value of *Q***

The significant rates of change of eccentricities introduced by this mechanism are surprising, because one might not have anticipated that subtle changes in a planetary spin rate, with consequent small changes in planetary oblateness, could play such a major role in orbital evolution, and even on the evolution of a second, distant planet. A key parameter is the choice of the value of $Q$, or equivalently the time lag $\Delta t$.

While CBL's adoption of the value $\Delta t = 200$ sec was useful for demonstrating the pumping mechanism, is that value actually plausible for such a system? The geophysical parameters adopted by CBL were selected to be those of Saturn. The time lag of 200 sec was said to be equivalent to a widely accepted value of $Q$ for Saturn: $3 \times 10^4$. However, the latter value was based on constraints from the orbit of Mimas (Goldreich and Soter 1966). The relevant frequency in that case is the difference between the spin rate of Saturn and the mean motion of Mimas, which is about 0.06 cycles/hour. According to the relation $\Delta t = 1/(\text{frequency} \times Q)$, the time lag for tides raised by Mimas was about 0.3 sec. CBL's formulation was based on an assumption that $\Delta t$ is independent of frequency. Therefore, for self-consistency, a value of $\Delta t = 0.3$ sec would be more appropriate than their adopted value of $\Delta t = 200$ sec. In that case, the SDT effect would be weakened by factor of ~600.

Given that the value $\Delta t = 200$ sec was adopted, we can ask what is the corresponding value of $Q$ in CBL's hypothetical system. The tidal period is approximately the mean motion ($n = 50$ rad/yr), so $Q = 1/(n\,\Delta t) = 3 \times 10^3$. Whether such a value is consistent with a Saturn-like planet driven at this low frequency is unknown, given the uncertainties regarding the processes of tidal dissipation described, for example, in **Section 3.2.1** above. However, if this $Q$ value (along with the corresponding $\Delta t = 200$ sec) does apply and the planet is considered to be Saturn-like, then the assumption that $\Delta t$ is independent of frequency is untenable. So the higher-order-eccentricity terms in CBL's formulation would no longer be valid.

Another way to look at the issue is that, if one's model has $\Delta t$ independent of frequency and equal to 200 sec, then this planet cannot be like Saturn. Even so, for purposes of constructing a hypothetical system to demonstrate the SDT effect, a planet with such properties cannot be ruled out. To address what other types of systems might have a significant SDT effect, we next compare it with the well-known direct tidal damping of the inner planet's eccentricity.

**4.2 Comparison with tidal damping of eccentricity**

The standard expression for the damping of an orbiting planet's eccentricity due to tides raised on the planet is (e.g. Goldreich and Soter 1966)

$$\frac{\dot{e}}{e} = -\frac{21}{2} n \frac{k}{Q} \frac{M}{m} (R/a)^5 \qquad (22)$$

For comparison, the pumping effect on the dominant eigenmode component of the inner planet, from Eq. (21c) can be expressed as

$$\frac{\dot{e}_{1p}}{e_{1p}} = 18 \frac{A_{21} F'_m F_m F_p}{S} e_{2m}^2 n^2 \frac{k^2}{\xi Q} \left(\frac{M}{m}\right)^2 (R/a)^8 \qquad (23)$$

Here we have assumed that the dissipation is small enough that $A < nS$ (**Section 2.2**), *i.e.* we approximate $\delta/(1+\delta^2)$ with $1/\delta$. Taking the ratio of the right sides of Eqs. (22) and (23) yields the ratio of the SDT pumping rate to the direct tidal damping rate (comparing absolute values of these rates):

$$\mathfrak{R}_{pd} \equiv \text{pumping/damping} = \frac{36}{21} \frac{A_{21} F'_m F_m F_p}{S} e_{2m}^2 nk\xi^{-1} \frac{M}{m} (R/a)^3 \qquad (24)$$

Typically $e_{2m}$ is the dominant eigenmode component for the outer planet, so $e_{2m} \approx e_2$. If the right side of Eq. (24) is comparable to or greater than 1, the SDT effect may be significant.

Eq. (24) can be used to test for the importance of the SDT effect in any particular real or hypothetical two-planet system. The right-hand side is dependent only on the mass values and orbital parameters of the system, and on the ratio $k/\xi$, which for our purposes is roughly ~1 for any planet. The quantity $A_{21} F'_m F_m F_p/S$ on the right-hand side is, according to the definitions in **Section 2**, a function of the components of the matrix $A_{ij}$, which in turn is a function of the masses and semimajor axes of the system. Thus evaluation of the $A_{ij}$ matrix, and insertion into Eq. (24), shows whether tidal effects are likely to increase or decrease the inner planet's eccentricity.

The expression for the ratio of the strength of SDT pumping relative to tidal damping may be even more useful if we use some additional approximations to put it more directly in terms of physical parameters, without the need to calculate the $A_{ij}$ values. Algebraic manipulation shows that the quantity $A_{21} F'_m F_m F_p/S$ has the form

$$\frac{A_{21} F'_m F_m F_p}{S} = \frac{A_{11}}{8 A_{21}^2} f(A_{12} A_{21}/A_{11}^2, A_{22}/A_{11}) \qquad (25)$$

Inspection of the functional form of $f$ as plotted in **Figure 3** shows that, if $A_{22}/A_{11} < 0.3$ and $A_{12} A_{21}/A_{11}^2 < 1$, then

$$f \approx 10 \, (A_{12} A_{21}/A_{11}^2)^2. \qquad (26)$$

Is it reasonable to assume that those two ratios are small enough to justify using Eq. (26)? Recall that $A_{11}$ and $A_{22}$ represent the apsidal precession rates for each of the two planets, without the effect of the eccentricity of the other. If we momentarily ignore any effects on precession other than that of the other planet (i.e. ignore GR and effects of the shape of the planets and star) then both $A_{22}/A_{11}$ and $A_{12} A_{21}/A_{11}^2$, are $\sim m_1 n_2 / m_2 n_1$. Thus, if the outer planet is more massive than the inner one, the conditions are satisfied for use of Eq. (26). If the inner planet's precession $A_{11}$ is increased significantly by GR or planetary shape, the simple functional form of $f$ in Eq. (26) would still be justified even if $m_2$ is somewhat less than $m_1$.

Assuming that the conditions are met for using Eq. (26) (*e.g.*, $m_2 > m_1$), Eq. (24) becomes

$$\mathfrak{R}_{pd} \approx 2 \, (A_{12}^2/A_{11}^3) \, e_{2m}^2 \, n \, (k/\xi) \, (M/m_1) \, (R/a_1)^3 \qquad (27)$$

The quadrupole approximations for $A_{12}$ and $A_{11}$ allow us to express these quantities in terms of masses and semimajor axes (e.g. Lee and Peale 2003, and CBL). Those expressions are most accurate for small values of $a_2/a_1$, but even for $a_2/a_1$ approaching 1, they give a fairly accurate value for the ratio $A_{12}/A_{11}$. They give a less accurate value for $A_{11}$, although it is still acceptable for our order-of-magnitude estimate of the pumping/damping ratio, as long as $a_2/a_1 < 0.7$. (If $a_2/a_1 \sim 0.6$ or $0.7$, the quadrupole approximation underestimates $A_{11}$ by a factor of 2 or 4 respectively.) If $a_1$ is closer to $a_2$, this approximation breaks down; on the other hand, in that regime, the theory should also be extended to include tides related to the outer planet as well as the inner planet. Inserting the quadrupole approximations (e.g. from CBL) for $A_{ij}$, which account only for the mutual effects of the planets, we find the simple formula

$$\mathfrak{R}_{pd} \approx 4\, e_{2m}^2\, (k/\xi)\, (M/m_1)\, (M/m_2)\, (a_2/a_1)(R/a_1)^3 \qquad (28)$$

This equation can be used to quickly screen systems for the possibility that the SDT effect is significant. For the hypothetical system of CBL, this equation gives $\mathfrak{R}_{pd} \approx 4$. However, in that system the effect of GR (not included in Eq. 28) would be an apsidal precession that roughly doubles $A_{11}$. Because $\mathfrak{R}_{pd} \propto A_{11}^{-3}$, its value is reduced to $\sim 0.5$. Thus our formulation would predict a negligible SDT effect relative to standard tidal damping. However, we have already seen that that our analytic approach is not applicable to cases with large eccentricities.

The strength of all tidal effects increases the closer the inner planet of a system is to its star. In general, tides do not become important for orbital evolution unless $a_1 < 0.1$ AU (e.g. Jackson et al. 2008). Decreasing $a_1$ would also increase the pumping/damping ratio according to Eq. (28). However, as in the hypothetical system of CBL, if $a_1$ is too small, GR increases $A_{11}$ and reduces $\mathfrak{R}_{pd}$. In other words, with decreasing $a_1$, tidal effects would be stronger, but the ratio of SDT pumping to tidal damping would decrease.

The ratio of the GR part of $A_{11}$ to the secular part (e.g. Ragozzine and Wolf 2009) is

$$\text{GR/secular} \equiv \mathfrak{R}_{GS} \approx 4\, (n_1 a_1/c)^2\, (M/m_2)\, (a_2/a_1)^3 \qquad (29)$$

If this ratio $\mathfrak{R}_{GS}$ is $<1$, then it has only a small effect on $\mathfrak{R}_{pd}$. More generally, the expression for $\mathfrak{R}_{pd}$ in Eq. (28) should contain a correction factor for GR:

$$\mathfrak{R}_{pd} \approx 4\, e_{2m}^2\, (k/\xi)\, (M/m_1)\, (M/m_2)\, (a_2/a_1)(R/a_1)^3 (1+ \mathfrak{R}_{GS})^{-3} \qquad (30)$$

This GR factor is not needed for the form of $\mathfrak{R}_{pd}$ given in Eq. (24) or (27) because GR is already included in $A_{11}$. Note too that for a planet close enough to its star that GR plays a role ($\mathfrak{R}_{GS} > 1$), $\mathfrak{R}_{pd}$ in Eq. (30) is proportional to $a_1^8$. Thus in this regime, as $a_1$ is reduced, tidal damping (which is $\propto a_1^{-5}$) becomes greater, but SDT is reduced. In other words, the SDT effect is unlikely to play a role if the inner planet is significantly affected by GR.

For very close-in planets, other effects contribute to the precession as well (e.g. Ragozzine and Wolf 2009). The ratio of the effect on $A_{11}$ of the prolate shape of the tidally deformed planet to that of GR is

$$\text{Tidal elongation/GR} \equiv \mathfrak{R}_{EG} \approx (k/3)\,(R/a_1)^5\,(c/n_1 a_1)^2\,(M/m_1), \tag{31}$$

so for very small $a_1$ the elongation plays an even greater role than either GR or the outer planet in determining the ratio of SDT pumping to tidal damping of $e_1$. This role can be taken into account by replacing $\mathfrak{R}_{GS}$ with $\mathfrak{R}_{GS}(1+\mathfrak{R}_{EG})$ in Eq. (30). However, this close in to the star, GR generally already precludes SDT from being significant according to Eq. (30).

It is important to emphasize that the expression (30) for the approximate ratio of SDT pumping to tidal damping includes several restrictions: (a) The values of $A_{22}/A_{11}$ and $A_{12}A_{21}/A_{11}^2$ must be sufficiently small (as they would be if $m_1 < m_2$); otherwise Eq. (24) should be used. (b) the value of $\delta = nS/A$ must be $\gg 1$; otherwise a correction factor $\delta^2/(1+\delta^2)$ must be applied. (c) the eccentricities must be small. Moreover, following CBL, we have considered tidal damping due to dissipation in the inner planet only. The additional eccentricity damping due to tides raised on the star would need to be taken into account to confirm whether or not SDT pumping would be adequate to increase the value of $e_1$. For that purpose, we note that the ratio of the damping rate due to tides on the star to the rate due tides on the planet is $\sim (m/M)^{1/3}(\rho/\rho_*)^{5/3}(Q/Q_*)$.

### 4.3 Where might tidal pumping be important?

If $\mathfrak{R}_{pd}$ is of the order of unity or greater, SDT pumping may exceed eccentricity damping. Thus evaluating it with the formulae above provides a way to identify two-planet systems where the spin-driven tidal effect may be important. The simpler form of Eq. (30) can be used for any case where the inner planet is less massive than the outer one, while Eq. (24) applies more generally.

Our formulation defines the region in parameter space where the effect should be taken into account. Unfortunately, visualizing the boundaries of that multi-dimensional space is not simple. A way to get a sense of the range of types of systems for which SDT pumping exceeds conventional tidal damping is to consider a hypothetical system where Eq. (30) has a value $\sim 1$, and then consider how changes in each parameter would affect the magnitude of the SDT effect.

Consider an Earth-like inner planet (with $k/Q \sim 1/20$ and Earth's mass and radius) at 0.08 AU, and an outer planet of 6 Earth masses at $a_2 = 1.5 a_1$ with $e_{2m} = 0.2$, orbiting a solar-mass star. In this case, the standard tidal damping rate for $e_1$ would be about $-2 e_1$/Gyr (Eq. 22); tides on the star would be negligible. Evaluation of Eq. (29) shows GR is small ($\mathfrak{R}_{GS} \sim 0.2$), but still large enough to reduce Eq. 30 by a factor of 2. Evaluating Eq. (30) shows that $\mathfrak{R}_{pd} \sim 1$, which would suggest that SDT pumping of $e_1$ could be comparable to the direct tidal damping. However, in this case $\delta = 0.27$, violating the assumption that it is large. Without that approximation, the correct value of $\mathfrak{R}_{pd}$ is reduced by a factor of $\delta^2/(1+\delta^2) = 15$: the SDT effect is an order of magnitude weaker than damping. Moreover, adjusting any of the parameters of the system fails to increase $\mathfrak{R}_{pd}$ to any significant degree.

For example, inspection of Eqs. (30) and (29) also shows that the stellar mass $M$ cannot be changed much from the solar value without significantly reducing $\mathfrak{R}_{pd}$ relative to the reference case. However, the weakening might be somewhat ameliorated if $a_1$ were different as well. That line of reasoning suggests consideration of a planet in the habitable zone of a low-mass star. Consider a star with M~0.1 solar masses. Its "habitable zone" (~ 0.02-0.04 AU) could be close enough to the star for tides to be important (Kopparapu et al. 2013). Eq. (30) suggests that the following planetary system would be a candidate for a significant SDT effect: A planet at $a_1$ = 0.03 AU with Earth's mass and radius, $k$=0.3, $\xi$=0.2, and $Q$=80; and an outer planet with a mass six times greater at $a_2$ = 0.16 AU with $e_{2m} = e_2 = 0.2$. In this case, $\Re_{pd}$ as evaluated by Eq. (30) is ~1. However, once again we find that $\delta$ is small; accordingly $\Re_{pd}$ is reduced to only 0.03. For this system, direct tidal damping is rapid; the timescale is only ~300 Myr. The SDT effect only reduces the damping rate by $2.8 \times 10^{-3}$/Gyr according to our analytical formulation. Numerical integration of CBL's equations give the same result. For all the cases we have considered, the SDT effect has only a very minor impact on the tidal evolution.

## 5. Discussion and Conclusion

Our analytic description of the spin-driven tidal effect identified by CBL confirms that this mechanism tends to pump the eccentricity of the inner planet in a two-planet system. It helps elucidate the basic physics of the process and confirms the interpretation by CBL, in which the lag in the response of the spin rate to the periodically changing eccentricity feeds back into the secular interaction by modifying the orbital precession rate.

An advantage of the analytic solution is that it shows the dependence of the mechanism on the various parameters of a two-planet system. Consideration of several examples suggests that the magnitude of the effect may be small for most realistic systems. The hypothetical case considered by CBL for illustrative purposes does show rapid pumping of the inner planet's eccentricity, but that result was obtained by assuming a large time lag in the tidal response. Whether that lag time and the equivalent value of $Q$ are justified is uncertain. As discussed in **Section 4.1**, the rationale that the adopted lag time is based on estimates of Saturn's $Q$ is inconsistent with the assumption that the lag time is independent of frequency. Nevertheless, given the complexity of real materials and planetary structure, and the consequent uncertainty regarding any planet's response to tidal distortion, no particular value of $Q$ can be definitively ruled out. Hence there may well be circumstances under which the SDT effect has played a role in shaping the current architecture of planetary systems.

In deriving our analytic solution, a number of approximations have been required. Strictly speaking, our formulae are accurate only for small eccentricities, because we have invoked classical secular theory. For CBL's hypothetical system, with $e$~0.5, our formula underestimates the strength of the SDT effect by a factor of ten. However, where $e$~0.2 or less, it appears to be reasonably accurate, according to comparison with numerical integration and with CBL's complementary analytic estimate.

Even formulations that include higher-order terms in eccentricities may not be as accurate as they appear, because they must depend strongly on a specific tidal model. The high-order terms are sensitive to the frequency dependence of the tidal response of a planetary body to a tidal potential, but the actual character of that response is unknown. The actual frequency dependence may be quite complex or otherwise very different from what has often been assumed (e.g. Ogilvie 2009, Efroimsky and Makarov 2013, Ferraz Mello 2013). Moreover it may not be amenable to the conventional Fourier separation and linear treatment of tidal components (Greenberg 2009). Because such gross uncertainty about the tidal response of real planets remains, analyses that include high-order terms in *e* may be precise, but not necessarily accurate representations of real planetary behavior. Thus the reduced precision of our formula for cases with large eccentricities may be commensurate with the underlying uncertainty about how tides really behave.

For any given system, the equations in **Section 4.2** allow an estimate of the strength of the eccentricity pumping relative to the usual tidal damping expected when the secular interactions with another planet are ignored. Equation (30) provides a convenient form, but several approximations were made in the simplification of the ratio $\Re_{pd}$ into the forms in Eqs. (28) and (30) from the more general form in Eq. (24). We assumed that both $A_{22}/A_{11}$ and $A_{12}A_{21}/A_{11}^2$ are small, for which having a less massive inner planet is a sufficient condition. More generally, the function *f* should be evaluated directly from Eq. (25), rather than using the simplified expression in Eq. (26). It is also important to recall that the expressions for $\Re_{pd}$ in **Section 4.2** assume that $A < nS$; otherwise, appropriate corrections should be made as demonstrated in **Section 4.3**.

The ratio $\Re_{pd}$ developed here can be used as a screening tool for identifying systems where the SDT effect should be considered, because it provides a comparison with the direct tidal damping due to dissipation in the inner planet. Given the approximations involved, such screening should be followed up with more precise numerical integration for any candidate system. In addition, any complete model of the evolution of a system should take into account other effects that affect the eccentricity, including for example tides on the star and the outer planet.

Although we have considered only two-planet systems, the general mechanism should apply even with systems with more planets. Moreover, we can generalize even further. The SDT effect works because a component of the orbital precession of one of the planets (due to its oblateness contributing to $A_{11}$) depends on the eccentricity, with a lagging response. Similar effects could be generated, for example, by the tidal bulge on a planet, or by the tidal bulge or the oblateness of the star. As the inventory of diverse planetary systems expands, and understanding of planetary tidal responses improves, the SDT effect may prove to play a significant role in the evolution of some systems.

## Acknowledgments

We thank Gwenaël Boué, Alexandre Correia, and an anonymous referee for helpful comments on the manuscript. Christa Van Laerhoven's work is supported by a NASA Earth and Space Science Fellowship. This project has been supported in part by Rory Barnes' NSF grant AST-110882.

**Figures**

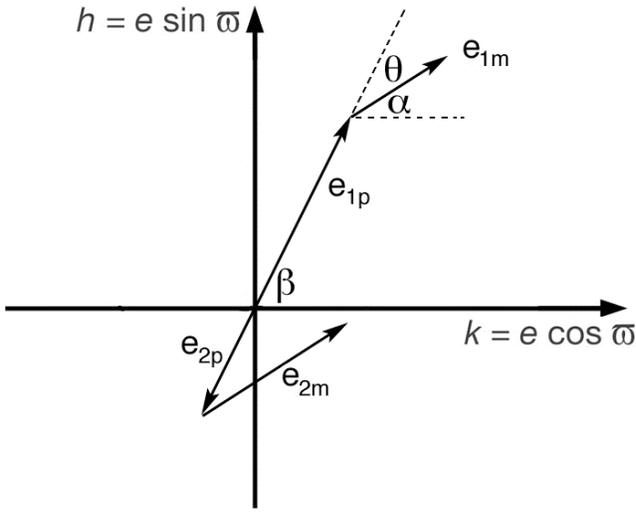

FIGURE 1: Each planet's eccentricity is a vector sum of contributions from each eigenmode (modes m and p). Contributions from modes m and p are aligned or antialigned, respectively. The directions given by $\alpha$ and $\beta$ rotate at rates given by the eigenfrequencies $g_m$ and $g_p$. Their relative rotation rate is $d\theta/dt = S = g_p - g_m$.

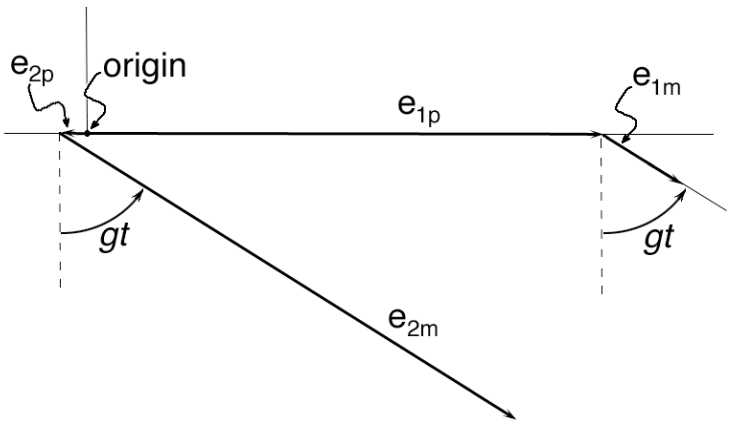

FIGURE 2: The solution of Equations (1) for the hypothetical case considered by Correia et al., showing the secular eccentricity behavior in a format similar to Figure 1, except here the horizontal axis is aligned with the mode p solution. The lengths of the component vectors represent the numerical values in the solution. The inner planet's eccentricity $e_1$ is the vector sum of $e_{1p}$ and $e_{1m}$, while $e_2$ is the vector sum of $e_{2p}$ and $e_{2m}$. Note that, for this system, $e_1$ is dominated by mode p, and $e_2$ is dominated by mode m. Thus $e_1$ oscillates about the value $e_{1p}$, and $e_2$ oscillates about the value $e_{2m}$. The values from our solution are $e_{1p} = 0.367$; $e_{1m} = 0.067$; $e_{2p} = 0.022$; $e_{2m} = 0.378$.

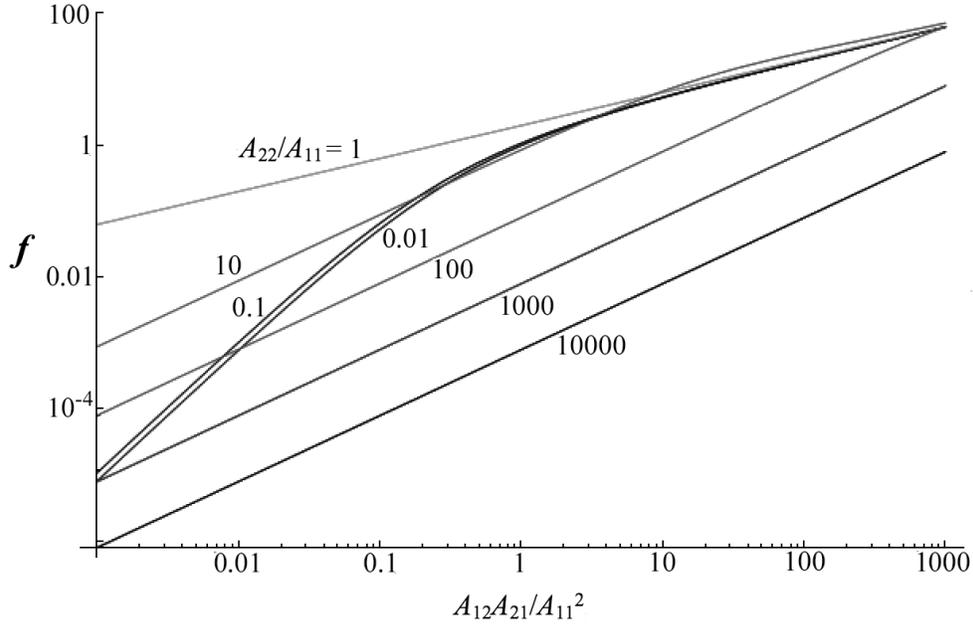

FIGURE 3: The function $f(A_{12}A_{21}/A_{11}^2, A_{22}/A_{11}) = 8A_{21}^3 F'_m F_m F_p / SA_{11})$ as defined in Eq. (25), showing that, if $A_{22}/A_{11} < 0.3$ and $A_{12}A_{21}/A_{11}^2 < 1$, then $f \approx 10\,(A_{12}A_{21}/A_{11}^2)^2$ as in Eq. (26).